\documentclass[preprint,nofootinbib]{revtex4}%
\usepackage{amssymb}
\usepackage{amsfonts}
\usepackage{amsmath}
\usepackage{amsmath}
\usepackage{graphicx}
\usepackage{hyperref}
\hypersetup{colorlinks,linkcolor={blue},citecolor={blue},urlcolor={black}} 

\usepackage[usenames]{color}%
\providecommand{\U}[1]{\protect\rule{.1in}{.1in}}
\providecommand{\U}[1]{\protect\rule{.1in}{.1in}}
\definecolor{blue}{rgb}{0,0,1}

\definecolor{red}{rgb}{1,0,0}

\begin{document}
\title{Slowly rotating and the accelerating $\alpha'$-corrected black holes in four and higher dimensions}
\author{Felipe Agurto-Sepúlveda$^{1,2}$, Mariano Chernicoff$^{3,4}$, Gaston Giribet$^5$, Julio Oliva$^2$ and Marcelo Oyarzo$^2$}
 
\affiliation{$^1$Departamento de Astronomía, Universidad de Concepción, Casilla 160-C, Chile;
}
\affiliation{$^2$Departamento de F\'{\i}sica, Universidad de Concepci\'on, Casilla, 160-C, Concepci\'on, Chile.}
\affiliation{$^3$Departamento de F\'{\i}sica, Facultad de Ciencias, Universidad Nacional Aut\'{o}noma de M\'{e}xico\\   A.P. 70-542, CDMX 04510, M\'{e}xico;}
\affiliation{$^4$Departament de F\'isica Qu\`antica i Astrof\'isica and Institut de Ci\`encies del Cosmos (ICC), \\ Universitat de Barcelona, {\it Mart\' i Franqu\`es 1, ES-08028, Barcelona, Spain.}}
\affiliation{$^5$Departamento de F\'{\i}sica, Universidad de Buenos Aires\\ Ciudad Universitaria, pabell\'on 1, 1428, Buenos Aires, Argentina;}

\begin{abstract}
We consider the low-energy effective action of string theory at order $\alpha '$, including $R^2$-corrections to the Einstein-Hilbert gravitational action and non-trivial dilaton coupling. By means of a convenient field redefinition, we manage to express the theory in a frame that enables us to solve its field equations analytically and perturbatively in $\alpha ' $ for a static spherically symmetric ansatz in an arbitrary number of dimensions. The set of solutions we obtain is compatible with asymptotically flat geometries exhibiting a regular event horizon at which the dilaton is well-behaved. For the 4-dimensional case, we also derive the stationary black hole configuration at first order in $\alpha '$ and in the slowly rotating approximation. This yields string theory modifications to the Kerr geometry, including terms of the form $a$, $a^2$, $\alpha '$ and $a\alpha '$. {In addition, we obtain the first $\alpha'$ correction to the C-metrics, which accommodates accelerating black holes. We work in the string frame and discuss the connection to the Einstein frame, for which rotating black holes have already been obtained in the literature.}
\end{abstract}
\maketitle

\section{Introduction}

Higher-curvature corrections to Einstein’s general relativity (GR) are ubiquitous in any sensible approach to quantum gravity, and they are a solid prediction of string theory \cite{GreenSchwarzWitten}. Even before the formulation of the latter, effective actions containing higher order contractions of the Riemann tensor were known to emerge in quantum field theory on curved spacetime \cite{BD} and in the semiclassical approach to quantum gravity. Such actions are the natural generalization of Einstein-Hilbert action, thus correcting GR in the ultraviolet (UV) regime. Also, from the mathematical point of view it was early understood that higher-curvature terms were natural in higher dimensions \cite{Lanczos, Lovelock, Lovelock2}; and, on general grounds, it is widely accepted that any attempt to formulate a sensible UV-complete theory will involve higher-curvature corrections in a way or another. In 1976, Stelle argued that gravitational actions which include terms quadratic in the curvature tensor are renormalizable \cite{Stelle}. This is due to the fact that non-linear renormalization of the graviton and the ghost fields suffices to absorb the non-gauge-invariant divergences that might arise. Stelle explained how these and other divergences may be eliminated in a way that simplifies the renormalization procedure, even when matter fields are coupled. Nevertheless, renormalizability is not the only issue: The inclusion of quadratic-curvature terms in the gravitational action typically introduces massive local degrees of freedom, apart from the massless graviton of GR \cite{Stelle2}. These extra modes organize themselves as a massive spin-2 and a massive spin-0 excitations, yielding a total of 8 local degrees of freedom. The massive spin-2 part of the field has negative energy, and this is the reason why it is usually asserted that, with exception of a few remarkable cases \cite{Lovelock, Lovelock2, Critical, Critical2}, augmenting the Einstein-Hilbert action with a finite set of higher-curvature terms yields ghosts when the theory is expanded about maximally symmetric vacua. The observation of \cite{Stelle2} motivated, in the early 80s, the search for ghost-free higher-curvature theories and consistent UV completions. Since then, actions containing higher-curvature terms were considered in the context of cosmology \cite{Starobinsky}, black hole physics \cite{BoulwareDeser}, and string theory \cite{Zwiebach}. In 1985, Zwiebach studied the compatibility between the presence of curvature squared terms and the absence of ghost modes in the low energy limit of string theory \cite{Zwiebach}. He argued that the so-called Einstein-Gauss-Bonnet (EGB) action was a good candidate for string effective action as it yields a ghost-free non-trivial gravitational self-interacting theory in any dimension greater than four, $d>4$. The EGB action is made out of a dimensionally extended version of the quadratic Chern-Gauss-Bonnet topological invariant, which, while being dynamically trivial in $d\leq 4$, does yield a UV correction to GR in $d>4$ with a single massless spin-2 excitation and with field equations of second order. The latter property makes EGB theory free of Ostrogradski instabilities. Still, in \cite{BoulwareDeser} Boulware and Deser showed that the EGB model, proposed in \cite{Zwiebach} as a stringy action, contains, in addition to flat spacetime, a second non-perturbative anti-de Sitter (AdS) vacuum which turns out to be unstable due to the presence of ghosts. This is nothing but the fact that actions that are polynomials of degree $k$ in contractions of the Riemann tensor generically yield $k$ different vacua, many of them being artifacts of the truncation of the effective theory. In \cite{BoulwareDeser2} the authors noticed that the inclusion of the dilaton field in the EGB effective action suffices to remove the spurious (A)dS vacuum permitted in its absence. They also showed that the spherically symmetric static solutions to the dilatonic EGB theory might have a well-defined asymptotic behaviour, being non-trivial, and being compatible with the existence of a regular event horizon at which the dilaton is well-behaved\footnote{Higher-curvature black holes were also studied in the context of thermodynamics \cite{Jacobson1} and many other subjects, like the holography \cite{Buchel2}, the weak-gravity conjecture \cite{Motl}, among others. For related early works on this paper, see \cite{Moura1, Moura2, Callan, Myersfoot, Cheung1, Cheung2, Cheung3}.} This was later confirmed by explicit examples, and here we will also provide a concrete realization of it. 

Soon after \cite{Zwiebach}, in a foundational paper of string theory \cite{GrossWitten}, Gross and Witten finally proved that the gravitational field equations of string theory actually contain higher-curvature corrections to GR. More precisely, they derived the modifications of the classical gravitational equations for the type II string theory by studying tree-level gravitational scattering amplitudes, and they determined the effective gravitational action up to quartic order in the curvature tensor, which corresponds to order $\mathcal{O}(\alpha’^3)$ string corrections. Unlike bosonic string theory, type II superstring theory in $d=10$ dimensions does not contain quadratic corrections to GR, and the cubic ones can be set to zero by fields redefinition -- although quadratic corrections can actually appear in Calabi-Yau compactification of the quartic actions, with the moduli playing the role of the couplings cf. \cite{Ferrara1, Ferrara2, Guica}--. In contrast, quadratic corrections do appear in critical bosonic and heterotic string theories. They were studied in \cite{Metsaev, Metsaev2} by Metsaev and Tseytlin, who checked the equivalence of the string equations of motion and the $\sigma$-model Weyl invariance conditions at order $\mathcal{O}(\alpha ’ )$. They obtained the functional dependence on the dilaton, the graviton, and the antisymmetric tensor. To do so, they first determined the $\mathcal{O}(\alpha ’ )$ terms in the string effective action starting from the expressions for the 3- and 4-point string scattering amplitudes; then, they computed the 2-loop $\beta$-function in the worldsheet $\sigma $-model. This results in an effective gravity action with quadratic-curvature ($R^2$) corrections coupled to the other massless fields of the theory; see also the important works \cite{Bergshoeff1, Bergshoeff2}, and for modern developments on $\alpha ' $ corrections in relation to $T$-duality and Double Field Theory see \cite{Hohm1, Hohm2, Hohm3, Hohm4, Hohm5, Hohm6, Hohm7, Marques1, Marques2, Marques3, Marques4} and references therein and thereof.

In recent years, with the advent of AdS/CFT correspondence and its ramifications, higher-curvature terms were reconsidered in the context of holography and the interest on them was revived. Probably the best-known example of this is the discussion of the higher-curvature terms in relation to the Kovtun-Starinets-Son (KSS) viscosity bound \cite{Brigante, Brigante0}, which showed that, for a class of conformal field theories (CFT) with a gravity dual with the EGB action, the shear viscosity to entropy density ratio could violate the conjectured KSS lower bound. This proved that the presence of higher-curvature terms could result in qualitatively new phenomena; see also \cite{Edelstein1, Edelstein2}. Microcausality violation in the CFT was also studied in the same type of scenario \cite{Brigante}, which was rapidly interpreted as evidence supporting the idea of a universal lower bound on the shear viscosity to entropy density ratio for all consistent theories. This triggered a long series of works devoted to check the consistency conditions of effective theories with higher-curvature modifications. For example, in \cite{Buchel} the authors discussed causality conditions in $R^2$ theories; they study causality violation in holographic hydrodynamics focusing on the EGB theory as a working example. In the latter theory, the value of the only $R^2$ coupling constant is related to the difference between the two central charges of the dual 4-dimensional CFT, and the authors of \cite{Buchel} showed that, when such difference is sufficiently large, causality is violated. This problem was also studied in \cite{Hofman}, where the author discussed the relation between causality constraints in the bulk theory and the condition of energy positivity in the dual CFT. He specifically argued that special care is needed when solving the classical equations of motion in the higher-curvature gravity theory, for which the study of causality problems may be subtle. Holography in presence of EGB gravity actions have been further studied in \cite{Buchel2} and in references thereof. The authors of \cite{Buchel2} studied the problem in arbitrary number of dimensions $d$ and established a holographic dictionary that relates the couplings of the gravitational theory to the universal numbers in the correlators of the stress tensor of the dual CFT, cf. \cite{BuchelMyers}. This allowed the authors to examine constraints on the gravitational couplings by demanding consistency of the CFT, and this yielded a much more general set of causality constraints. 

Both in the context of AdS/CFT and in other scenarios, the consistency conditions for higher-curvature theories were intensively studied in the last fifteen years. This line of research has continued and a much more general picture of the set of consistency conditions has been accomplished. Causality, locality, stability, hyperbolicity and other aspects were revisited. In \cite{Gruzinov}, it was shown how causality constrains the sign of the stringy $R^4$ corrections to the Einstein-Hilbert action, giving a general restriction on candidate theories of quantum gravity. In \cite{Gomberoff}, a special type of pathology that the truncated EGB theory exhibits was studied. This is a phase transition driven by non-perturvative effects that might take place in gravitational theories whenever higher-curvature corrections with no extra fields are considered. In \cite{Camanho}, Maldacena et al. studied causality constraints on corrections to the graviton 3-point coupling. They considered higher-curvature corrections to the graviton vertex in a weakly coupled gravity theory and they derived stringent causality constraints. By considering high energy scattering processes, they noticed a potential causality violation that might occur whenever additional Lorentz invariant structures are included in the graviton 3-point vertex. They argued that such a violation could be cured by the addition of an infinite tower of extra massive higher-spin fields such as those predicted by string theory. This problem was later reconsidered by many authors, cf. \cite{Reall}.

Motivated by this renewed interest in higher-curvature gravity, in the last years there have been important developments in the subject, and many new higher-curvature models were proposed and studied. The list includes the quasitopological theories \cite{quasitopological, quasitopological2, quasitopological3}, the critical gravity theories in AdS \cite{Critical, Critical2}, the so-called Einsteinian cubic gravity \cite{Cubic, Cubic2}, and their generalizations \cite{quasitopologicalinfty, quasitopologicalinfty2}. Black holes have recently been studied in all these setups \cite{quasitopological2, Chernicoff, CubicBH, Grandi}, as well as in string theory inspired scenarios \cite{Cano1, Cano2, Cano3}; see also \cite{Maeda2009, Giribeto, Ohta} and references therein and thereof. Here, we will present and study analytic, static, spherically symmetric solutions to the $\alpha ' $-corrected gravity action in arbitrary dimension $d$ and including a non-vanishing dilaton coupling. We will consider the graviton-dilaton sector of the low-energy effective action of string theory with $R^2$ terms in a specific frame that will enable us to solve the problem explicitly to order $\mathcal{O}(\alpha ' )$ in the entire spacetime. Our solutions manifestly show that the theory is compatible with static, spherically symmetric solutions which are asymptotically flat  and exhibit a regular event horizon at which the dilaton is well-behaved. The paper is organized as follows: In section II, we present the gravity theory in a convenient frame. We briefly discuss the field redefinition ambiguity to the relevant order, and we use it to solve the adequate ansatz. The field equations are written down and solved, and the black hole solution for $d=4$ is presented. In section III, we study the black hole thermodynamics. This amounts to work out the Wald entropy formula, which, as usual in this type of setup, yields corrections to the Bekenstein-Hawking area law. The mass of the solution may then be inferred from the first law of black hole mechanics. In section IV, we perform a consistency check of the previous formulae by explicitly computing the black hole mass by means of the Iyer-Wald method for conserved charges, which shows perfect agreement. We also show the agreement with the Euclidean action approach. In section V, we generalize our result by introducing angular momentum in the slowly rotating approximation. We derive a stationary metric that represents stringy modifications to the Kerr geometry. {In section VI, we obtained the $\alpha'$ correction to the C-metric, which accommodates accelerating black holes. While we work in the string frame, in section VII we discuss the frame transformation that maps out theory to the Einstein frame, including the higher curvature corrections. In the latter frame, rotating solutions have already been studied in the literature, and we discuss the precise relation between the two frames.} In section VIII, we generalize the static solution by presenting the explicit form of the dilatonic black hole solution in arbitrary dimension $d$.

\section{Dilatonic black hole}

We consider the low-energy effective action of string theory including $\alpha^{\prime}$ corrections to the graviton-dilaton sector; namely \cite{Metsaev, Metsaev2}
\begin{equation}
I\left[  g_{\mu\nu},\phi\right]  =\int_M d^{d}x\sqrt{-g}e^{-2\phi}\left[
R+4\left(  \nabla\phi\right)  ^{2}+\alpha\ R_{\mu\nu\lambda\rho}R^{\mu
\nu\lambda\rho}+\mathcal{O}\left(\alpha^2\right)\right]  \ ,
\end{equation}
where we denoted $\alpha=\frac{1}{8}\alpha^{\prime}$. We are not considering the dependence on the $B$-field here. Performing field redefinition $g_{\mu\nu}\rightarrow g_{\mu\nu}+\delta g_{\mu\nu}$, $\phi\rightarrow\phi+\delta\phi$ with
\begin{eqnarray}
&\delta\phi=-\frac{\alpha}{2}\left(R+4(2d-5)\partial_{\mu}\phi\partial^{\mu}\phi\right)\, ,\\
&\delta g_{\mu\nu}=-4\alpha\left(R_{\mu\nu}-4\partial_\mu\phi\partial_\nu\phi+4g_{\mu\nu}\partial_\alpha\phi\partial^\alpha\phi\right)\, ,
\end{eqnarray}
one obtains the action in a frame that is convenient for the computation we want to undertake; namely
\begin{align}
I\left[  g_{\mu\nu},\phi\right]    & =\int d^{d}x\sqrt{-g}e^{-2\phi}\left[
R+4\left(  \nabla\phi\right)  ^{2}\right.  \nonumber\\
& \left.  +\alpha\left(  R_{\mu\nu\lambda\rho}R^{\mu\nu\lambda\rho}-4R_{\mu
\nu}R^{\mu\nu}+R^{2}-16\left(  \partial_{\mu}\phi\partial^{\mu}\phi\right)
^{2}\right)  +\mathcal{O}\left(  \alpha^{2}\right)  \right]  \ ,\label{action}%
\end{align}
up to six-derivative operators of order $\mathcal{O}\left(  \alpha^{\prime2}\right)$, cf. \cite{Maeda2011}. As the $R^2$-terms take the form of the $4d$ Euler characteristic, the field equations of the theory in this frame are of second order in an explicit manner. Let us consider first the case in $d=4$. The field equations derived from (\ref{action}) are given by
\begin{eqnarray}
&&G_{\mu \nu }+4\partial _{\mu }\phi \partial _{\nu }\phi -2g_{\mu \nu
}\partial _{\rho }\phi \partial ^{\rho }\phi +2S_{\mu \nu }-2g_{\mu \nu
}S_{\ \rho }^{\rho }+\alpha H_{\mu \nu }\left. =\right. 0\ , \\
&&R+4\partial _{\rho }\phi \partial ^{\rho }\phi +4S_{\ \mu }^{\mu }+\alpha
L_{GB}-32\alpha \left( \nabla ^{\mu }\left( \partial _{\rho }\phi \partial
^{\rho }\phi \right) \partial _{\mu }\phi +\left( \partial _{\rho }\phi
\partial ^{\rho }\phi \right) S_{\ \mu }^{\mu }+\frac{1}{2}\left( \partial
_{\rho }\phi \partial ^{\rho }\phi \right) ^{2}\right) \left. =\right. 0\ , 
\notag \\
&&
\end{eqnarray}%
where $G_{\mu \nu }=R_{\mu \nu }-\frac 12 Rg_{\mu \nu }$ is the Einstein tensor, and where
\begin{eqnarray}
S_{\rho \sigma } \equiv e^{2\phi }\nabla _{\rho }\left( e^{-2\phi }\nabla
_{\sigma }\phi \right) \ , \ \
L_{GB} \equiv R^{\mu \nu \rho \sigma }R_{\mu \nu \rho \sigma }-4R^{\mu \nu
}R_{\mu \nu }+R^{2}\ ,
\end{eqnarray}%
and%
\begin{eqnarray}
H_{\mu \nu } &=&S_{\mu \nu }R-4S_{\ (\mu }^{\sigma }R_{\nu )\sigma }+2S_{\
\sigma }^{\sigma }R_{\mu \nu }+2S^{\sigma \lambda }R_{\mu \sigma \lambda \nu
}-8\left( \partial _{\rho }\phi \partial ^{\rho }\phi \right) \partial _{\mu
}\phi \partial _{\nu }\phi  \\
&&+g_{\mu \nu }\left( 2\left( \partial _{\rho }\phi \partial ^{\rho }\phi
\right) ^{2}-S_{\ \sigma }^{\sigma }R+2S_{\ \lambda }^{\sigma }R_{\ \ \sigma
}^{\lambda }\right) \ .  \notag
\end{eqnarray}
Lagrangian $L_{GB}$ is the integrand of the $4$-dimensional Chern-Gauss-Bonnet topological invariant which, in the absence of the dilaton and in $d=4$, yields the Euler characteristic; this is the EGB quadratic gravity Lagrangian.

We are interested in solving the equations above for a static spherically symmetric spacetime with non-trivial dilaton profile (later, in Section V, we will generalize the solution to the stationary, non-static case). In order to do so, we work perturbatively at order $\mathcal{O}(\alpha )$, and propose the ansatz
\begin{align}
\phi\left(  r\right)    & =\phi_{0}+\alpha\phi_{1}\left(  r\right)  \ ,\\
ds^{2}  & =-\left(  1+\alpha N_{1}\left(  r\right)  \right)  ^{2}\left(
1-\frac{\mu}{r}+\alpha f_{1}\left(  r\right)  \right)  dt^{2}+\frac{dr^{2}%
}{1-\frac{\mu}{r}+\alpha f_{1}\left(  r\right)  }+r^{2}d\Omega^{2}\ , \label{corrected metric N}
\end{align}
where $\phi_1(r)$, $N_1(r)$, and $f_1(r)$ are functions of the radial coordinate $r$ to be determined; $\mu $ is an arbitrary constant; $d\Omega ^2$ is the constant-curvature metric on the unit sphere. The solution we will find in this way will be valid up to order $\mathcal{O}(\alpha)$. Plugging this ansatz in the field equations and expanding up to first order in $\alpha$, we obtain a remarkably simple system of equations which lead to the following general
solution%
\begin{equation}
\phi\left(  r\right)  =\phi_{0}+\alpha\left(  A+B\log\left(  \frac{r-\mu}%
{r}\right)  -\frac{2}{\mu r}-\frac{1}{r^{2}}-\frac{2\mu}{3r^{3}}\right)\label{Dil}
\end{equation}
with $A$ and $B$ being two arbitrary constants; the former constant appears merely as a shift of $\phi_{0}$ which does not enter in the metric, and so it can be absorbed by redefining $\bar{\phi}_0=\phi_0+\alpha A$, which gives the value that the dilaton takes at infinity; notice that, at infinity, (\ref{Dil}) goes like $\phi\simeq \bar{\phi}_0 +\mathcal{O}(1/r)$. Up to $\mathcal{O}(\alpha)$ terms, for the metric we find
\begin{align}
g^{rr}  & =1-\frac{\mu}{r}+\alpha\left(  -\frac{\mu B}{r}\log\left(
\frac{r-\mu}{r}\right)  +\frac{C}{r}+\frac{2}{r^{2}}+\frac{\mu}{r^{3}}%
-\frac{10}{3}\frac{\mu^{2}}{r^{4}}\right)  \ ,\nonumber \\
g_{tt}  & =\frac{\mu}{r}-1-\alpha\left(  \frac{B\left(  2r-3\mu \right)  }%
{r}\log\left(  \frac{r-\mu}{r}\right)  +D+\frac{4}{r^{2}}+\frac{5\mu}{3r^{3}%
}+\frac{2\mu^{2}}{r^{4}}-\frac{\mu^{2}D-\mu C+2\mu^{2}B+8}{\mu \, r}\right) \nonumber
\end{align}
where $D$ and $C$ are other two integration constants. The former can be eliminated by rescaling the time coordinate as $t\to t/({1+\alpha D})$.

If we define $
r_{+}=\mu+\alpha\mu_{1}$, we can easily find the $\alpha $-corrected location of the event horizon by solving for $\mu_{1}$ as a function of the integration constants. This amounts to demand $g^{rr}\left(  r_{+}\right)  =g_{tt}\left(  r_{+}\right)
=0$, which is actually required for the horizon to be regular. Expanding up to first order in $\alpha$, this yields
\begin{align}
\left(  B\log\left(  \frac{\mu}{\alpha\mu_{1}}\right)  +\frac{\mu_{1}+C}{\mu
}-\frac{1}{3\mu^{2}}\right)  \alpha+O\left(  \alpha^{2}\right)    & =0\ ,\\
\left(  B\log\left(  \frac{\mu}{\alpha\mu_{1}}\right)  -2B+\frac{\mu_{1}+C}%
{\mu}-\frac{1}{3\mu^{2}}\right)  \alpha+O\left(  \alpha^{2}\right)    &
=0\ ,
\end{align}
from which we consequently obtain that $B=0$. Therefore, the $\alpha^{\prime}%
$-corrected black hole configuration reads%
\begin{align}
g^{rr} (r) & =1-\frac{\mu}{r}+\alpha\left(  \frac{C}{r}+\frac{2}{r^{2}}+\frac
{\mu}{r^{3}}-\frac{10}{3}\frac{\mu^{2}}{r^{4}}\right)  +\mathcal{O}(\alpha ^2)\ ,\\
g_{tt} (r) & =\frac{\mu}{r}-1-\alpha\left( \frac{4}{r^{2}}+\frac{5\mu}%
{3r^{3}}+\frac{2\mu^{2}}{r^{4}}-\frac{8-\mu C}{\mu \, r}\right) \label{gtt} +\mathcal{O}(\alpha ^2)   \ , \\
{\phi}(r)    & =\bar{\phi}_{0}-\alpha\left(  \frac{2}{\mu r}+\frac
{1}{r^{2}}+\frac{2\mu}{3r^{3}}\right) +\mathcal{O}(\alpha ^2) \ ,
\end{align}
and, up to $\mathcal{O}(\alpha)$ corrections, the location of the horizon is
\begin{equation}
r_{+}=\mu+\alpha\left(  \frac{1}{3\mu}-C\right)  \, . \label{corrected horizon}
\end{equation}

The solution we have just derived is asymptotically flat, and it exhibits a smooth event horizon at $r_+$, where the dilaton remains finite:
\begin{equation}
\phi(r_+)=\bar{\phi}_0 -\frac{11}{3}\frac{\alpha}{ r_+^{2}}+\mathcal{O}(\alpha ^2)\, ; \ \ \ \phi(\infty )=\bar{\phi}_0 +\mathcal{O}(\alpha ^2).
\end{equation}

In the following sections, we will analyze the physical properties of this solution, we will compute its conserved charges, and, finally, we will generalize it to $d\geq 4$ dimensions.

\section{Thermodynamics}

The thermodynamics of higher-curvature black holes has been studied for a long time \cite{Jacobson1, Jacobson2, Jacobson3}, and in a vast number of contexts. Here, we will focus on the properties of the black hole solution we just presented. The Wald formula gives the entropy as a Noether charge computed at the horizon. This is given by the following integral on the horizon $\mathcal{%
H}$%
\begin{equation}
S=\frac{\beta}{4} \int_{\mathcal{H}}\, \sqrt{-g}\, \epsilon _{\mu \nu
\rho \sigma }\, q^{\mu \nu }dx^{\rho }\wedge dx^{\sigma },  \label{S wald}
\end{equation}%
in $d=4$ spacetime dimensions, with $\beta $ being the periodicity of the Euclidean
time. The Noether pre-potential associated to this charge is given by%
\begin{equation}
q^{\mu \nu }\equiv -2\left( E^{\mu \nu \rho \sigma }\nabla _{\rho }\xi
_{\sigma }+2\xi _{\rho }\nabla _{\sigma }E^{\mu \nu \rho \sigma }\right) 
\end{equation}%
and 
\begin{equation}
E_{\ \ \rho \sigma }^{\mu \nu }\equiv \frac{\partial \mathcal{L}}{\partial
R_{\ \ \mu \nu }^{\rho \sigma }}\ .  \label{Etensor}
\end{equation}%
For the action (\ref{action}), the tensor (\ref{Etensor}) and the
Noether pre-potential take the following form
\begin{eqnarray}
E_{\ \ \rho \sigma }^{\mu \nu } &=&\frac{1}{2}e^{-2\phi }\left( \delta
_{\rho \sigma }^{\mu \nu }+\alpha \delta _{\rho \, \sigma \, \nu _{3}\nu
_{4}}^{\mu \, \nu \, \mu _{3}\mu _{4}}R_{\ \ \ \ \mu _{3}\mu _{4}}^{\nu
_{3}\nu _{4}}\right) +\mathcal{O}\left( \alpha ^{2}\right) \ , \\
q^{\mu \nu } &=&2e^{-2\phi }T^{\mu \nu }+\alpha e^{-2\phi }\left( 4T^{\mu
\nu }R+16T^{\sigma \lbrack \mu }R_{\ \sigma }^{\nu ]}+4T^{\rho \sigma }R_{\
\ \rho \sigma }^{\mu \nu }\right) +\mathcal{O}\left( \alpha ^{2}\right) \ ,
\end{eqnarray}
respectively, where we have defined $T^{\rho \sigma }\equiv 4\xi ^{\lbrack \rho
}\nabla ^{\sigma ]}\phi -\nabla ^{\lbrack \rho }\xi ^{\sigma ]}$. Evaluating the Wald entropy (\ref{S wald}) for our solution, we obtain
\begin{equation}
S=16\pi ^{2}e^{-2\bar{\phi } _{0}}\mu ^{2}-32\pi ^{2}\alpha \, e^{-2\bar{\phi }_{0}}\left(
C\mu -8\right) +\mathcal{O}\left( \alpha ^{2}\right) \ .\label{entropy1}
\end{equation}%
If we naturally identify
\begin{equation}
e^{-2 \bar{\phi } _{0}}= \frac{1}{16\pi G}\ ,
\end{equation}%
$G$ being the $4d$ Newton constant, the leading term in (\ref{entropy1}) reproduces the Bekenstein-Hawking entropy, while the order $\mathcal{O}(\alpha)$ terms yield corrections to it. More precisely, we find
\begin{eqnarray}
S =\frac{\pi \mu ^{2} }{G}+\frac{16\pi \alpha }{G}\left(1-\frac 18  C\mu
\right) +\mathcal{O}\left( \alpha ^{2}\right)\, .\label{linearinmu}
\end{eqnarray}%
Notice that (\ref{linearinmu}) depends both on $\mu$ and $C$. The dependence of $C$ can be traced back to the fact that $\alpha(8-C\mu)/\mu$ is the $\mathcal{O}(\alpha)$ correction to the parameter in front of the Newtonian piece $\sim 1/r$ in the component $g_{tt}$ of the metric, cf. (\ref{gtt}). Then, using (\ref{corrected horizon}), the entropy can also be written as
\begin{eqnarray}
S =\frac{\pi r_{+}^{2}}{G}+\frac{46\pi \alpha}{3G}  +\mathcal{O}\left( \alpha ^{2}\right) \, . \label{Daentropy}
\end{eqnarray}
Notice that the potential term linear in $r_+$ (i.e. the one that could come from the term linear in $\mu $ in (\ref{linearinmu})) has cancelled out. In fact, at order $\mathcal{O}(\alpha )$, in virtue of the field equations, the computation reduces to that of the full action evaluated on the undeformed GR solution $f_1=N_1=\phi_1=0$. This means that, at that order, the only correction to the area law $S =\frac{A}{4G}$ is given by a positive constant. On the same grounds, corrections of the form $\mathcal{O}(\alpha r_+^{d-4}/G)$ are expected in higher dimensions.

Next, let us compute the Hawking temperature. We can do this by resorting to the Euclidean formalism. However, it is convenient to first simplify the expressions a bit. We can write $\mu$ as a function of $r_+$ by simply inverting (\ref{corrected horizon}), which yields
\begin{eqnarray}
g_{tt}(r) &=&-1+\frac{1}{r}\left( r_{+}+\frac{23\alpha }{3r_{+}}\right)-\alpha
\left( \frac{4}{r^{2}}+\frac{5r_{+}}{3r^{3}}+\frac{2r_{+}^{2}}{r^{4}}\right)
+\mathcal{O}\left( \alpha ^{2}\right) \ , \label{last_configuration} \\
g^{rr}(r) &=&1-\frac{1}{r}\left( r_{+}-\frac{\alpha }{3r_{+}}\right) +\alpha
\left( \frac{2}{r^{2}}+\frac{r_{+}}{r^{3}}-\frac{10r_{+}^{2}}{3r^{4}}\right)
+\mathcal{O}\left( \alpha ^{2}\right) \ , \label{last_configuration2} \\
\phi (r)  &=&\bar{\phi } _{0}-\alpha \left( \frac{2}{r_{+}r}+\frac{1}{%
r^{2}}+\frac{2r_{+}}{3r^{3}}\right) +\mathcal{O}\left( \alpha ^{2}\right) \ . \label{last_configuration3}
\end{eqnarray}
This gives the periodicity condition for the real section of the Euclidean geometry to be regular at $r=r_+$; namely
\begin{equation}
\beta =4\pi r_{+}+\frac{44\pi \alpha }{3r_{+}}+\mathcal{O}\left( \alpha
^{2}\right) \ , \label{beta_last_configuration}
\end{equation}
which results in the black hole temperature
\begin{equation}
T=\frac{1}{4\pi r_{+}} \left( 1 -\frac{11}{3}\frac{\alpha }{r_+^2}\right)  +\mathcal{O}\left(
\alpha ^{2}\right)\, . \label{LaT}
\end{equation}
This corrects the Hawking formula for GR at scales $r_+ \simeq \alpha^{1/2}$. This result, together with the expression (\ref{Daentropy}) for the entropy, yields the first law type relation
\begin{equation}
\delta E \equiv T\delta S =\delta \left( \frac{ r_{+} }{2G}+\frac{11\alpha }{%
6Gr_{+}}\right) +\mathcal{O}\left(
\alpha ^{2}\right)\, ,
\end{equation}
from which, up to subleading orders in $\alpha $, we can obtain the gravitational energy
\begin{equation}
E-E^{(0)}= \frac{ r_{+} }{2G} \left( 1+\frac{11}{3}\frac{\alpha }{r_+^2} \right) \label{masita}
\end{equation}
$E_0$ being an integration constant that corresponds to the energy of the reference background. Below, we will confirm this result by rederiving the gravitational energy using the Iyer-Wald method for computing Noether charges. It is also worth noticing that, if we insist in extrapolating the formulae above for small values of $r_+$, which is not well justified as higher-order terms are expected to be relevant in that regime, then the formula obtained for the specific heat changes its sign and becomes positive within the range $\frac{11}{3}\alpha  <r_+^2 < {11}\alpha $; the black hole temperature (\ref{LaT}) vanishes at the lower bound $r_+^2=\frac{11}{3}\alpha $.

\section{Conserved charges}

In order to compute the gravitational energy of the solution, we have to supplement the bulk action with the appropriated boundary terms. In the case of the higher curvature action \eqref{action}, the boundary term to be added reads 
\begin{align}
    I_{BT}\equiv \int_{\partial M} d^3x\sqrt{-h}e^{-2\phi}\left[ 2 K +4\alpha \delta^{\mu_1 \mu_2 \mu_3} _{\nu_1 \nu_2 \nu_3} K^{\nu_1}_{ \mu_1} \left(  \frac{1}{2} \mathcal{R}^{\nu_2 \nu_3}_{ \ \ \mu_2 \mu_3 }- \frac{1}{3}K^{\nu_2}_{ \mu_2}K^{\nu_3}_{ \mu_3} \right) \right] \equiv \int_{\partial M} d^3x\sqrt{-h} \mathcal{B}\label{TerminoB}
\end{align}
where $K$ is the trace of the extrinsic curvature $K^{\mu}_{\nu}$, and $\mathcal{R}^{\mu \nu}_{\ \rho \sigma }$ and $h_{\mu\nu }$ are the intrinsic curvature and the induced metric on $\partial M$, respectively; cf. \cite{Nathalie}. The contribution (\ref{TerminoB}) renders the variational principle well-posed. Then, the energy of the spacetime, which corresponds to the black hole mass, is given by the following integral on the sphere at infinity, $S^2_\infty$; namely
\begin{align}
    M=\int_{S^2_\infty}\left( \boldsymbol{Q}[\boldsymbol{t}]-\boldsymbol{t}\cdot  \boldsymbol{B}\right) \ , \label{energy}
\end{align}
where $\boldsymbol{Q}[\boldsymbol{t}]$ is the Hodge dual of the Noether pre-potential for the killing vector $\boldsymbol{t}=\partial_t$ and
\begin{equation}
    \boldsymbol{B}=\frac{1}{3!}\,\mathcal{B}\, n^{\sigma}\,\sqrt{-g}\,\epsilon_{\mu \nu \rho \sigma} \,dx^{\mu}\wedge dx^{\nu} \wedge dx^{\rho} \ .
\end{equation}
In flat spacetime, the trace of the extrinsic curvature is ${K}^{(0)}=\frac 2r$ which gives a divergent piece in the action principle as the volume element contributes with $r^2$. To obtain a finite action principle and a finite energy definition, we have to subtract the extrinsic curvature of flat spacetime to each piece of the extrinsic curvature appearing in the formulae above. In other words, we have to define
\begin{equation*}
    \bar{K}_{\mu \nu}\equiv K_{\mu \nu}-{K}^{(0)}_{\mu \nu}\, ,
\end{equation*}
using flat space as a background reference; this corresponds to set $E^{(0)} =0$ for Minkowski spacetime. According to this, the energy content of the spacetime, as defined in \eqref{energy}, precisely gives \begin{equation}
    M=\frac{r_+}{2G}\left( 1+\frac{11}{3}\frac{\alpha}{r_+^2} \right)  \, ,
\end{equation}
which agrees with (\ref{masita}).

Another crosscheck for this result can be done by means of the Euclidean action formalism. In the saddle point approximation, the on-shell Euclidean action gives the partition function; namely 
\begin{equation}
    \log {Z}\simeq I^E+I^E_{BT} \, , \label{logZ}
\end{equation}
where the superscript $E$ stands for Euclidean. It is worth emphasizing that, at order $\mathcal{O}(\alpha )$, the computation of the Euclidean action reduces to the evaluation of the full action $I^E+I^E_{BT}$ on the undeformed GR solution. Therefore, the energy of the configuration can be simply derived from (\ref{logZ}) by computing
\begin{equation}
    \bar{E} =- \frac{\partial \log {Z}}{\partial \beta } \ . \label{energy_Z}
\end{equation}
The on-shell action computed with $\bar{K}_{\mu \nu}$ for the configuration (\ref{last_configuration})-(\ref{last_configuration3}) with the Euclidean time periodicity \eqref{beta_last_configuration} turns out to be finite, and it reads
\begin{eqnarray}
I^E+I^E_{BT}=-\frac{\pi r_{+}^2}{G}-\frac{10 \alpha \pi }{3 G}\, .
\end{eqnarray}
From this expression, we easily find
\begin{equation}
    \bar{E} = \frac{r_+}{2G} \left(  1+\frac{11}{3}\frac{\alpha }{r_+^2}\right) \, ,\label{LaE}
\end{equation}
which, again, exactly reproduces (\ref{masita}) at the right order. This results in an $\mathcal{O}(\alpha )$, $r_+$-dependent correction to the GR Smarr formula; namely
\begin{equation}
    T\, S\, - \, \frac 12\, \bar{E} \, = \, \frac{2\alpha }{Gr_+}\, .
\end{equation}
At order $\mathcal{O}(\alpha)$ this is equivalent to an additive constant in the entropy.

\section{Adding angular momentum}

Black hole solution (\ref{corrected metric N}) can be generalized to the stationary non-static case, and the analytic expression in the slowly rotating approximation can also be found following the similar perturbative method as before. At first order in $\alpha$ and including the rotation parameter in linear and quadratic terms as well as in terms of the form $a\alpha$, the solutions reads%
\begin{eqnarray}
ds^{2} & =&-\left(1-\frac{\mu }{r}+\frac{\mu a^{2}\cos^{2}\theta}{r^{3}}+\alpha
f_{1}\left(  r\right)  \right)dt^{2}+2a\left(  -\frac{\mu \sin^{2}\theta}{r}+\alpha \, h_{t\varphi}\left(
r,\theta\right)  \right)  dtd\varphi\, +\nonumber \\
&&\ \ \ \ \  \left(  \frac{1}{1-\frac{\mu }{r}+\alpha
g_{1}\left(  r\right)  }-\frac{\left(  \left(  \mu -r\right)  \cos^{2}%
\theta+2r\right)  a^{2}}{\left(  r-\mu \right)  ^{2}r}\right)  dr^{2}\, +\label{rota1} \\
&&  \ \ \ \left(  r^{2}+a^{2}\cos^{2}\theta\right)  d\theta^{2}+\left(  \left(
r^{2}+a^{2}\right)  \sin^2\theta+\frac{a^{2}\mu \sin^{4}\theta}{r}\right)
d\varphi^{2}\, ,\nonumber
\end{eqnarray}
with%
\begin{align}
f_{1}\left(  r\right)   &  =\frac{2\mu^{2}}{r^{4}}+\frac{5\mu }{3r^{3}}+\frac
{4}{r^{2}}-\frac{8}{\mu r}\ ,\\
g_{1}\left(  r\right)   &  =-\frac{40}{3r^{4}}+\frac{\mu }{r^{3}}+\frac{2}%
{r^{2}}\ ,\\
h_{t\varphi}\left(  r\right)   &  =\sin^{2}\theta\left(  \frac{\hat{C}}{r}%
+\frac{2\mu ^{2}+3\mu r+6r^{2}}{r^{4}}\right)\, ,\label{rota2}
\end{align}
and with $\hat{C}$ being a new integration constant that, at this order, comes to renormalize the angular momentum; see (\ref{tecalculasteelJ2}) below. The scalar configuration is%
\begin{equation}
\phi\left(  r\right)  =\phi_{0}-\alpha\left(  \frac{2\mu }{3r^{3}}+\frac{1}%
{r^{2}}+\frac{2}{\mu r}\right)  \ .\label{rota3}
\end{equation}
One can verify that, expanding both in the Gauss-Bonnet coupling, $\alpha$,
and in the rotation parameter, $a$, all the field equations are solved at the right order; namely%
\begin{equation}
E_{\mu\nu}=\mathcal{O}\left(  \alpha a^{2}, \alpha^2\right)  \ .
\end{equation}
The angular momentum can be computed by using the Wald formalism, which yields a form
\begin{equation}
    J=-\int_{S^2_{\infty }} \boldsymbol{Q}[\partial_\varphi ] \label{tecalculasteelJ}
\end{equation}
with $\boldsymbol{Q}[\partial_\varphi ]$ representing the Hodge dual of the Noether pre-potential for the Killing vector $\partial_\varphi$. The angular momentum of the spacetime is given by
\begin{equation}
    J=\frac{a\mu}{2G}\left( 1-\frac{\alpha \hat{C}}{\mu}
 \right) \, \, .\label{tecalculasteelJ2}
\end{equation}
Solution (\ref{rota1})-(\ref{rota2}) gives string theory modification to Kerr geometry. In particular, we see order $\mathcal{O}(a\alpha )$ modifications to the off-diagonal term in the Boyer-Lindquist coordinates. This will result in deviations from the GR prediction of the Lense-Thirring precession. It will also induce modifications to the spheroidal shape of the shadow of a rotating black hole; see \cite{Eiroa} and references thereof.

\section{Accelerating black holes}

Let us consider the following ansatz for the metric and for the dilaton%
\begin{eqnarray}
ds^{2} &=&\frac{\Omega \left( x,y\right) }{A^{2}\left( x+y\right) ^{2}}%
\left( -F\left( y\right) dt^{2}+\frac{dy^{2}}{F\left( y\right) }+\frac{dx^{2}%
}{G\left( x\right) }+G\left( x\right) d\varphi ^{2}\right) \ ,
\label{Cmetric} \\
\phi  &=&\phi \left( x,y\right) \ .
\end{eqnarray}%
Assuming the expansion $\phi =\phi _{0}\left(x,y\right)+\alpha \phi _{1}\left( x,y\right) +%
\mathcal{O}(\alpha ^{2})$, $F=F_{0}\left( y\right) +\alpha F_{1}\left(
y\right) +\mathcal{O}(\alpha ^{2})$, $G=G_{0}\left( x\right) +\alpha
G_{1}\left( x\right) +\mathcal{O}(\alpha ^{2})$ and $\Omega \left(
x,y\right) =1+\alpha \omega \left( x,y\right) +\mathcal{O}(\alpha ^{2})$. In
General Relativity, the ansatz (\ref{Cmetric}), leads to the C-metric which
accommodates accelerating black holes (see \cite{Griffiths:2009dfa} for a modern interpretation as well as a historical review), even in the presence of minimally coupled, self interacting scalar fields \cite{Anabalon:2012ta}. Here $A$, stands for
the acceleration and $\Omega \left( x,y\right) =1$ in General Relativity in vacuum.

To the lowest order on the string tension $\alpha $, Einstein equations lead to
\begin{equation}
F\left( y\right) =F_{0}\left( y\right) =f_{3}y^{3}+f_{2}y^{2}+f_{1}y+f_{0}%
\text{ and }G\left( x\right) =G_{0}\left( x\right)
=f_{3}x^{3}-f_{2}x^{2}+f_{1}x-f_{0}\ ,  \label{GR}
\end{equation}%
fulfilling $G\left( \xi \right) =-F\left( -\xi \right) $. Here $f_{i}$ with $%
i=\{0,...,3\}$ are integration constants. The quadratic, the linear or the $%
f_{0}$ term in the polynomials (\ref{GR}) can be removed by a simultaneous, constant shift
of the independent variables $(x,y)$, maintaining the form of the metric (\ref%
{Cmetric}). For future purposes, it is better to keep all the $f_{i}$ as
non-vanishing at the moment.

The field equations of the $alpha'$-corrected theory \eqref{action}, at linear order in $\alpha$ are solved by%
\begin{eqnarray}
F_{1}\left( y\right)  &=&d_{3}y^{3}+d_{2}y^{2}+d_{1}y+d_{0}\ ,
\label{Fcorrection} \\
G_{1}(x) &=&f_{3}h_{1}x^{3}+3f_{3}h_{2}x^{2}+\frac{\left(
3d_{3}f_{1}-3f_{1}f_{3}h_{1}-6f_{2}f_{3}h_{2}+3d_{1}f_{3}-2d_{2}f_{2}\right) 
}{3f_{3}}x\\
&&-\frac{\left(
-6f_{0}f_{3}h_{1}-3f_{1}f_{3}h_{2}+3d_{0}f_{3}-d_{2}f_{1}+6d_{3}f_{0}\right) 
}{3f_{3}}\ ,  \label{Gcorrection} \\
\omega (x,y) &=&2\phi _{1}(x,y)+\frac{3f_{3}j_{1}x+3\left(
2f_{3}h_{1}+f_{3}j_{1}-2d_{3}\right) y-6f_{3}h_{2}-2d_{2}}{3\left(
x+y\right) f_{3}}\ ,  \label{omegacorrection}
\end{eqnarray}%
leading to the following inhomogenoeus, PDE for $\phi _{1}\left( x,y\right) $%
\begin{eqnarray}
0 &=&\left( x+y\right) \left( G_{0}\left( x\right) \frac{\partial ^{2}\phi
_{1}}{\partial x^{2}}+F_{0}\left( y\right) \frac{\partial ^{2}\phi _{1}}{%
\partial y^{2}}\right)   \nonumber \\
&&+(f_{3}x^{3}+3f_{3}x^{2}y-2f_{2}xy-f_{1}x+f_{1}y+2f_{0})\frac{\partial
\phi _{1}}{\partial x}\\
&&+(f_{3}y^{3}+3f_{3}xy^{2}+2f_{2}xy+f_{1}x-f_{1}y-2f_{0})\frac{\partial \phi
_{1}}{\partial y}-6A^{2}f_{3}^{2}\left( x+y\right) ^{5}\ .  \label{eqacc}
\end{eqnarray}
Here the constants $(d_i,f_j,h_k,j_l)$ are new integration constants that emerge from the integration of the field equations at linear order in $\alpha$. Even though the equation \eqref{eqacc} seems not to admit an analytic solution, it can be
solved as a power series in the acceleration $A$, around $A=0$. In order to
be able to take the limit $A=0$ in (\ref{Cmetric}), it is useful to perform
the following change of coordinates (see Chapter 14 of \cite{Griffiths:2009dfa}):
\begin{equation}
x=-\cos \theta \text{, }y=\frac{1}{Ar}\text{, }t=A\tau \ ,
\label{changecoords}
\end{equation}%
and choosing 
\begin{equation}
f_{2}=-f_{0}=1\text{ and }\ f_{3}=-f_{1}=-2mA\ ,  \label{solgr}
\end{equation}
leads to the following parametrization for the C-metric in General
Relativity%
\begin{equation}
ds_{0}^{2}=\frac{1}{\left( 1-Ar\cos \theta \right) ^{2}}\left( -Q_{0}\left(
r\right) d\tau ^{2}+\frac{dr^{2}}{Q_{0}\left( r\right) }+\frac{r^{2}d\theta
^{2}}{P_{0}\left( \theta \right) }+P_{0}\left( \theta \right) r^{2}\sin
^{2}\theta d\varphi ^{2}\right) \ ,
\end{equation}%
with%
\begin{eqnarray}
Q_{0}\left( r\right)  &=&\left( 1-\frac{2m}{r}\right) \left(
1-A^{2}r^{2}\right) \ ,  \label{Q0} \\
P_{0}\left( \theta \right)  &=&1-2mA\cos \theta \ .  \label{P0}
\end{eqnarray}%
In terms of $\left( r,\theta \right) $, and choosing the constants $%
f^{\prime }s$ as in (\ref{solgr}), the equation (\ref{eqacc}) is
integrated, order by order in the acceleration $A$. For such purpose, it is
convenient to choose%
\begin{equation}
\phi _{1}\left( r,\theta \right) =\left( 1-Ar\cos \theta \right) H\left(
r,\theta \right) \ ,  \label{escalar}
\end{equation}%
with%
\[
H\left( r,\theta \right) =\sum_{i=0} H_{i}\left( r,\theta \right)A^i 
\]%
which leads to the following functions at the lowest orders%
\begin{eqnarray}
H_{0}\left( r,\theta \right)  &=&-\frac{4m}{3r^{3}}-\frac{1}{r^{2}}-\frac{1}{%
mr}\ , \\
H_{1}\left( r,\theta \right)  &=&\left( \frac{26m}{3r^{2}}-\frac{1}{r}%
\right) \cos \theta \ , \\
H_{2}\left( r,\theta \right)  &=&\frac{2m\left( 2\sin ^{2}\theta -23\right) 
}{3r}\ .
\end{eqnarray}%
Other solutions are possible, but they lead to logarithmic or divergent
behavior for the dilaton as $r\rightarrow \infty $.

In order to clarify the meaning of the plethora of integration constants which remain
arbitrary on the metric functions, it is useful to reconstruct the full,
corrected spacetime \eqref{Cmetric}, in $\left( r,\theta \right) $ coordinates. The change
of coordinates (\ref{changecoords}), induces the presence of $A^{-1}$ terms
in the metric coming from the terms (\ref{Fcorrection})-(\ref%
{omegacorrection}) written in terms of $(r,\theta)$, which are removed by setting $d_{3}=0$.  Imposing the
absence of divergences at $\theta =0$ and $\theta =\pi $ on the metric
functions suffices to fix all the remaining integration constants, but $j_1$, leading to $d_{2}=d_{1}=d_{0}=h_{1}=h_{2}=0,$ which in consequence
leads to vanishing corrections of the function $F$ and $G$, namely%
\begin{equation}
F_{1}\left( r,\theta\right) =0,\ G_{1}(r,\theta)=0\ ,
\end{equation}%
and to a conformal factor $\omega \left( r,\theta\right) $ given by%
\begin{equation}
\omega (r,\theta)=2\phi _{1}(r,\theta)\ ,
\end{equation}%
where we have also set $j_{1}=0$ since a non-vanishing value of $j_1$ can be absorbed on the dilaton's additive, arbitrary constant $\phi_0$.

Putting all these ingredients together, lead to the corrected metric which is given by
\begin{equation}
ds^{2}=\frac{1+2\alpha \phi _{1}\left( r,\theta \right) }{\left( 1-Ar\cos
\theta \right) ^{2}}\left( -Q_{0}\left( r\right) d\tau ^{2}+\frac{dr^{2}}{%
Q_{0}\left( r\right) }+\frac{r^{2}d\theta ^{2}}{P_{0}\left( \theta \right) }%
+P_{0}\left( \theta \right) r^{2}\sin ^{2}\theta d\varphi ^{2}\right) \ .
\label{metricrtheta}
\end{equation}%
Here $\phi _{1}\left( r,\theta \right) $ is given by (\ref{escalar}) and $%
Q_{0}\left( r\right) $ and $P_{0}\left( \theta \right) $ given by (\ref{Q0})
and (\ref{P0}), respectively. One can check, that the metric (\ref%
{metricrtheta}), with $\phi _{1}\left( r,\theta \right) $ in (\ref{escalar}%
), solve the field equations of the theory \eqref{action}, disregarding terms of the
form $\mathcal{O}\left( \alpha ^{2}\right) $ and $\mathcal{O}\left( \alpha
A^{3}\right) $, i.e. when evaluated on the corrected C-metric, the field
equations vanish up to%
\begin{equation}
E_{\mu \nu }=\mathcal{O}\left( \alpha ^{2}\right) +\mathcal{O}\left( \alpha
A^{3}\right) \ .
\end{equation}%
It is very interesting to notice that the regularity
conditions lead us to move the whole effect of the $\alpha $-correction to
the conformal factor. The solution can be found to higher orders on the
acceleration by performing the integration of the PDE (\ref{eqacc}), at
the desired order on $A$, ofter moving to $\left( \tau ,r,\theta \right) $
coordinates via (\ref{changecoords}), in such a manner that the limit of
vanishing acceleration is regular. 

\section{Mapping to the Einstein frame}

Recently in \cite{Cano3}, the authors constructed the dimensional reduction of the Heterotic String on a flat torus, to dimension  four, and constructed rotating solutions, perturbativelly in the rotation parameter, including the first $\alpha'$-correction, in the Einstein frame. It is interesting to compare the setup we have considered here, defined by the action \eqref{action}, with the one of reference \cite{Cano3}, where disregarding the contribution of the $B_{\mu\nu}$-field, leads to an action of the form
\begin{equation}
I\left[g^{\prime}_{\mu\nu},\tilde{\phi}\right]=\int d^{4}x\sqrt{-g^{\prime }}\left( R^{\prime }-\frac{1}{4}\nabla
_{\mu }\tilde{\phi}\nabla _{\nu }\tilde{\phi}g^{\prime \mu \nu }+\alpha e^{-%
\tilde{\phi}}\left( R_{\ \ \ \rho \sigma }^{\prime \mu \nu }R_{\ \ \ \mu \nu
}^{\prime \rho \sigma }-4R_{\ \sigma }^{\prime \nu }R_{\ \nu }^{\prime
\sigma }+R^{\prime 2}\right) \right) \ .\label{actionP}
\end{equation}
Considering a Weyl transformation of the form
\begin{eqnarray}
g_{\mu \nu } &\mapsto &g_{\mu \nu }^{\prime }=e^{\Phi }g_{\mu \nu }\ ,
\end{eqnarray}%
where $\Phi$ is some scalar function on the spacetime, the transformation of the quadratic scalars constructed with the Riemann tensor are given by
\begin{eqnarray*}
R_{\ \ \ \rho \sigma }^{\prime \mu \nu }R_{\ \ \ \mu \nu }^{\prime \rho
\sigma } &=&e^{-2\Phi }\left[ R_{\ \ \ \rho \sigma }^{\mu \nu }R_{\ \ \ \mu
\nu }^{\rho \sigma }-4R_{\ \sigma }^{\nu }\nabla _{\nu }\Phi ^{\sigma
}+2R_{\ \rho }^{\nu }\Phi ^{\rho }\Phi _{\nu }-R\Phi ^{\lambda }\Phi
_{\lambda }\right.  \\
&&+D_{2}\nabla _{\sigma }\Phi ^{\nu }\nabla _{\nu }\Phi ^{\sigma }+\left(
\square \Phi \right) ^{2}-D_{2}\nabla _{\sigma }\Phi _{\nu }\Phi ^{\sigma
}\Phi ^{\nu }+D_{2}\square \Phi \Phi ^{\lambda }\Phi _{\lambda }\left. +%
\frac{1}{8}D_{2}D_{1}\left( \Phi ^{\lambda }\Phi _{\lambda }\right) ^{2}%
\right] \ , \\
R_{\ \sigma }^{\prime \nu }R_{\ \nu }^{\prime \sigma } &=&e^{-2\Phi }\left[
R_{\ \sigma }^{\nu }R_{\ \nu }^{\sigma }-D_{2}R^{\nu \sigma }\nabla _{\nu
}\Phi _{\sigma }-R\square \Phi +\frac{1}{2}D_{2}R^{\nu \sigma }\Phi _{\nu
}\Phi _{\sigma }-\frac{1}{2}D_{2}R\Phi ^{\lambda }\Phi _{\lambda }\right.  \\
&&+\frac{1}{4}D_{2}^{2}\nabla _{\sigma }\Phi ^{\nu }\nabla _{\nu }\Phi
^{\sigma }+\frac{1}{4}\left( 3D-4\right) \left( \square \Phi \right) ^{2}+%
\frac{1}{16}D_{2}^{2}D_{1}\left( \Phi ^{\lambda }\Phi _{\lambda }\right) ^{2}
\\
&&\left. -\frac{1}{4}D_{2}^{2}\nabla _{\sigma }\Phi _{\nu }\Phi ^{\sigma
}\Phi ^{\nu }+\frac{1}{4}D_{2}\left( 2D-3\right) \square \Phi \Phi ^{\lambda
}\Phi _{\lambda }\right] \ , \\
R^{\prime 2} &=&e^{-2\Phi }\left[ R^{2}-2D_{1}R\square \Phi -\frac{1}{2}%
D_{1}D_{2}R\left( \partial \Phi \right) ^{2}+D_{1}^{2}\left( \square \Phi
\right) ^{2}\right.\\
&&\left.+\frac{1}{2}D_{1}^{2}D_{2}\square \Phi \left( \partial \Phi
\right)^{2}+\frac{1}{16}D_{1}^{2}D_{2}^{2}\left( \partial \Phi \right)
^{4}\right] 
\end{eqnarray*}
where $\Phi_\lambda=\nabla_\lambda\Phi$ and $D_p:=(D-p)$. These expressions lead to the following transformation of the Gauss-Bonnet density, which we had actually worked out in arbitrary dimension $D$
\begin{eqnarray}
&&R_{\ \ \ \rho \sigma }^{\prime \mu \nu }R_{\ \ \ \mu \nu }^{\prime \rho
\sigma }-4R_{\ \sigma }^{\prime \nu }R_{\ \nu }^{\prime \sigma }+R^{\prime
2}\nonumber\\
&=&e^{-2\Phi }\left[ R_{\ \ \ \rho \sigma }^{\mu \nu }R_{\ \ \ \mu \nu
}^{\rho \sigma }-4R_{\ \sigma }^{\nu }R_{\ \nu }^{\sigma
}+R^{2}+4D_{3}G^{\nu \sigma }\nabla _{\nu }\Phi _{\sigma }-2D_{3}R^{\nu
\sigma }\Phi _{\nu }\Phi _{\sigma }\right.  \notag\\
&&-\frac{1}{2}D_{4}D_{3}R\left( \partial \Phi \right) ^{2}-D_{3}D_{2}\nabla
_{\sigma }\Phi ^{\nu }\nabla _{\nu }\Phi ^{\sigma }+D_{3}D_{2}\nabla
_{\sigma }\Phi _{\nu }\Phi ^{\sigma }\Phi ^{\nu }  \notag \\
&&+\left. D_{3}D_{2}\left( \square \Phi \right) ^{2}+\frac{1}{2}%
D_{3}^{2}D_{2}\square \Phi \left( \partial \Phi \right) ^{2}+\frac{1}{16}%
D_{4}D_{3}D_{2}D_{1}\left( \partial \Phi \right) ^{4}\right] \ ,  \notag \\
&=&e^{-2\Phi }\left( \mathcal{G+P}\right)  \label{GB primed}
\end{eqnarray}%
where 
\begin{equation}
\mathcal{G}=R_{\ \ \ \rho \sigma }^{\mu \nu }R_{\ \ \ \mu \nu }^{\rho \sigma
}-4R_{\ \sigma }^{\nu }R_{\ \nu }^{\sigma }+R^{2}
\end{equation}%
and $\mathcal{P}$ stands for the remaining terms. Replacing these expression in \eqref{actionP} and choosing the scalar $\Phi$ as
\begin{equation}
\Phi=-\frac{4}{D-2}\phi\ ,
\end{equation}
and identifying the scalar field in \eqref{actionP} $\tilde\phi$ as
\begin{equation}
\tilde\phi=\frac{4}{D-2}\phi ,
\end{equation}
and setting $D=4$, leads to
\begin{eqnarray}
&&I\left[g^{\prime}_{\mu\nu}=e^{-2\phi}g_{\mu\nu},\tilde{\phi}=2\phi\right]\notag \\
&=&\int d^{4}x\sqrt{-g}e^{-2\phi }\left[ R+4\left( \partial \phi \right)
^{2}\right. +\alpha \left. \left( \mathcal{G}-32\left( \partial \phi \right)
^{4}-16G^{\mu \nu }\nabla _{\mu }\phi \nabla _{\nu }\phi +24\square \phi
\left( \partial \phi \right) ^{2}\right) \right] \label{actionCtransf}
\end{eqnarray}
where we have disregarded boundary terms. Notice that this action belongs to the family of the most general String Theory actions $\alpha'$-corrected, which after field-redefinitions lead to second order field equations, since each of the derivative terms for the scalar sector belongs to the Horndeski family \cite{Horndeski:1974wa}. Indeed, upon comparison with eq. (2.6) of reference \cite{Maeda2011} one can read from the action \eqref{actionCtransf} that the coefficients $(\lambda,\mu,\nu)$ of reference \cite{Maeda2011} are given by
$\lambda =-32,\mu =-16$ and $\nu =24$, and they indeed fulfil the consistency constraint $\lambda +2\left( \mu +\nu \right) +16=0$. The relation among the relative coefficients of the higher derivative operators of the scalar attests about the UV finiteness of the action. In consequence, using the results of reference \cite{Maeda2011}, one can see that the action \eqref{actionP} of reference \cite{Cano3} in the Einstein frame and our action \eqref{action} in the string frame are related by a field redefinition, composed with a change of frame.

Therefore, our static and rotating solutions of Sections II and IV, correspond to a change of frame of the solutions found in \cite{Callan} and \cite{Cano3}, respectively, composed with a field redefinition. On the other hand, the solution corresponding to accelerating black holes we presented in Section VI, is completely new.

\section{The $d$-dimensional solution}

The analytic solution that we constructed in $d=4$ can be generalized to arbitrary
dimension $d$ in a similar manner, although, as we will see, the form of the general case is a bit more involved. 

Consider the ansatz
\[
ds^{2}=-\left(  1+\alpha N_{1}\left(  r\right)  \right)  ^{2}\left(  1-\left(
\frac{\mu}{r}\right)  ^{d-3}+\alpha f_{1}\left(  r\right)  \right)
dt^{2}+\frac{dr^{2}}{1-\left(  \frac{\mu}{r}\right)  ^{d-3}+\alpha
f_{1}\left(  r\right)  }+r^{2}d\Omega_{d-2}^{2}\ ,
\]
where now $d\Omega_{d-2}^{2}$ is the constant-curvature metric on the unit $(d-2)$-sphere. By plugging this ansatz in the field equations for $d$ generic, we find the following general solution
\begin{align*}
N_{1}(r)    & =-\frac{C_3}{\mu^2 f_{0}(r)} \left(\frac{\mu}{r}\right)^{d-3} +\frac{(d-3)(d-2)}{(d-1)\mu^2}\left(\frac{\mu}{r}\right)^{d-1} \left[ F(r)+\frac{1}{f_0(r)}\left( \left(\frac{\mu}{r}\right)^{d-3} d-1 \right) \right] \\
& +\frac{C_{3}}{(d-3)\mu^2}\log{f_{0}(r)} +\frac{C_2}{\mu^2}\ ,\\
f_{1}(r)    & =-\frac{(d-3)}{(d-1)\mu^2}\left(\frac{\mu}{r}\right)^{2d-4}\left[ (d-3)(d-2)F(r)+2(2d-3) \right] +\frac{C_1}{\mu^2}\left(\frac{\mu}{r}\right)^{d-3} \\
&-\frac{C_3}{\mu^2}\left(\frac{\mu}{r}\right)^{d-3} \log{f_0(r)}  \ ,
\end{align*}
and
\begin{align*}
\phi_{1}(r)    & =\frac{\left(  d-3\right)  \left(  d-2\right)^2
}{2\left(  d-1\right)  r^{2}}\left(  \frac{\mu}{r}\right)  ^{d-3}\left(
  F(r)  -1\right)   +\frac{(d-2)C_3}{2(d-3)\mu^2}\log{f_0(r)} +\frac{C_{4}}{\mu^{2}}
\end{align*}
where $f_{0}\left(  r\right)  =1-\left(  \frac{\mu}{r}\right)  ^{d-3}$ and where $F(r)$ is given in terms of the hypergeometric function,
\[
F\left(  r\right) \, = \, {}_{2}F_{1}\left(  1,\frac{d-1}%
{d-3},2\, \frac{  d-2  }{  d-3  },\left(  \frac{\mu}%
{r}\right)  ^{d-3}\right)\, ;
\]
$C_{1},\, C_{2},\, C_{3}$, and $C_{4}$ are integration
 constants, analogous to the constants $A$, $B$, $C$, and $D$ of the $d=4$ case; for example, one can identify $C_3=\mu^2 B +2$, $C_4= \mu^2\bar{\phi}_0$, and so on. Some of these constants can be fixed as in the 4-dimensional case, i.e. by rescaling time coordinate $t$, shifting the zero-mode of $\phi $, neglecting $\mathcal{O}(\alpha^2)$ remnants, imposing a globally flat asymptotic behavior as $r\rightarrow\infty$, and requiring regularity at the event horizon. One can easily check that the 4-dimensional solution studied in the previous sections is recovered in the case $d=4$. To see this, it is convenient to consider the relation
\begin{equation}
 {}_{2}F_{1}(1,3,4,z)=-\frac {3}{2z^3} \left[ z(z+2)+2 \log(1-z)\right]\, ,
\end{equation}
with $z=1-f_0(r)$. The presence of logarithmic terms $\log f_0(r)=\log (1-z)$ in the functions $N_1(r)$, $f_1(r)$ and $\phi_1(r)$ is related to the fact that the third argument of the hypergeometric function, $c=2\frac{(d-2)}{(d-3)}$, turns out to be an integer for some dimensions ($d=4,5$). The logarithm, which in any case tends to zero at large $r$, disappears if one chooses $C_3$ appropriately.

\section{Conclusions}
 
Summarizing, the solutions we have presented in this paper describe static, spherically symmetric configurations in the graviton-dilaton sector of the $d$-dimensional low-energy stringy effective action (\ref{action}). This includes square-curvature terms and a non-vanishing dilaton coupling. We have used the freedom of field redefinitions to recast the action in a form that leads to second order field equations, {while still working in the string frame}. The set of solutions includes asymptotically flat black holes with regular event horizons, which behave as thermodynamic objects, just like expected. As a working example, we first focused on the the 4-dimensional case, which is given by (\ref{last_configuration})-(\ref{last_configuration3}). We derived the corrections to the thermodynamic variables introduced by the higher-curvature effects; we computed the Bekenstein-Hawking entropy (\ref{Daentropy}), the Hawking temperature (\ref{LaT}), and the mass formula (\ref{LaE}) including the $\mathcal{O}(\alpha )$ effects. The computation of the Noether charges was shown to be in exact agreement with the first law of black hole mechanics as derived from the Wald entropy formula; the Euclidean action formalism also reproduces these results. {We also obtained the correction to the C-metric, which contains accelerating black holes. We have show that regularity conditions imply that the whole modifications is contained within the conformal factor of the spacetime.} We also integrated the equations of motion for the stationary, non-static solution in the slowly rotating approximation. This yields stringy corrections to the Kerr geometry in four dimensions. Although, in contrast to the 4-dimensional case, the field equations in arbitrary dimension $d$ are more involved, we showed that in the static case they can still be solved explicitly in terms of hypergeometric functions. {Static and rotating black holes in these theories have already been considered in the literature, when the theory is expressed in a different fashion, which is possible due to the freedom of field redefinition. Considering such freedom, we found the precise relation between our setup and the frames previously considered on the literature. In particular, we mapped our theory to the Einstein frame, including the higher curvature corrections, and we showed the equivalence of our setup and that of reference \cite{Cano3}, where rotating solutions were already presented.} These solutions may serve as working examples to investigate higher-curvature stringy effects in a concrete setup.

\section*{Acknowledgements}

We thank Cristóbal Corral and Nicolás Grandi for their enlightening comments.
We also thank the anonymous referee who pointed out the possibility of mapping our theory to a
frame where rotating solutions have already been constructed in Ref. [66], which led us to include Secs. VI
and VII. F. A. is thankful for funding via FONDECYT (Fondo Nacional de Desarrollo Científico y Tecnológico) REGULAR 1201280. The work of M. C. is partially supported by Mexico’s National Council of Science and Technology
(CONACyT) Grant A1-S-22886, DGAPA-UNAM Grant IN107520, and a PASPA fellowship. The work of G. G. was
supported by CONICET (Consejo Nacional de Investigaciones Científicas y Técnicas) and ANPCyT (Agencia Nacional de Promoción Científica y Tecnológica)
 through Grants 755 PIP1109-2017 and PICT-2019-00303. The work of M. O. is partially funded by Beca ANID (Agencia Nacional de Investigación y Desarrollo)
 de Doctorado Grant 21222264 and FONDECYT Grant 1221504. J. O.
thanks the support of proyecto FONDECYT REGULAR
1221504, Proyecto de Cooperación Internacional 2019/13231-7 FAPESP/ANID, and Dirección de Postgrado,
Universidad de Concepción, through Grant UCO1866.

\end{document}